\documentclass[11pt,letter]{article}
\pdfoutput=1
\usepackage{jheppub}
\usepackage[utf8]{inputenc}

\usepackage{color}
\usepackage[usenames,dvipsnames]{xcolor}
\usepackage{amsmath}
\usepackage{mathtools}
\usepackage{esint}
\usepackage{pifont}
\usepackage{bbold}

\usepackage{bbm}
\usepackage{verbatim}   
\usepackage{subfigure}
\usepackage{acronym}

\usepackage{amsfonts}
\usepackage{amssymb}
\usepackage{mathrsfs}
\usepackage{graphicx}
\usepackage{multirow}
 \usepackage{slashed}
 \usepackage{url}

\usepackage{caption}

\usepackage{soul}

\newcommand{\eehz}{e^+e^- \to hZ}

\title{Long Live the Higgs Factory: Higgs Decays to Long-Lived Particles at Future Lepton Colliders}

\author[a]{Samuel Alipour-Fard,}
\emailAdd{samuel\_alipourfard@ucsb.edu}
\author[a]{Nathaniel Craig,}
\emailAdd{ncraig@physics.ucsb.edu}
\author[a,b]{Minyuan Jiang,}
\emailAdd{dg1522023@smail.nju.edu.cn}
\author[a]{and Seth Koren}
\emailAdd{koren@physics.ucsb.edu}

\affiliation[a]{Department of Physics, University of California, Santa Barbara, CA 93106, USA}
\affiliation[b]{Department of Physics, Nanjing University, Nanjing 210098, P.R.C.}

\abstract{
We initiate the study of exotic Higgs decays to long-lived particles (LLPs) at proposed future lepton colliders, focusing on scenarios with displaced hadronic final states. Our analysis entails a realistic tracker-based search strategy involving the reconstruction of displaced secondary vertices and the imposition of selection cuts appropriate for eliminating the largest irreducible backgrounds. The projected sensitivity is broadly competitive with that of the LHC and potentially superior at lower LLP masses. In addition to forecasting branching ratio limits, which may be freely interpreted in a variety of model frameworks, we interpret our results in the parameter space of a Higgs portal Hidden Valley and various incarnations of neutral naturalness, illustrating the complementarity between direct searches for LLPs and precision Higgs coupling measurements at future lepton colliders.}

\begin{document}

\maketitle

\section{Introduction}

Following the discovery of the Higgs boson in 2012 \cite{Aad:2012tfa, Chatrchyan:2012xdj}, the precision study of its properties has rapidly become one of the centerpieces of the physics program at the LHC. The expansion of this program beyond the LHC has become one of the key motivators for proposed future accelerators, including lepton colliders such as CEPC \cite{CEPCStudyGroup:2018rmc, CEPCStudyGroup:2018ghi}, FCC-ee \cite{Gomez-Ceballos:2013zzn}, the ILC \cite{Behnke:2013xla, Baer:2013cma}, and CLIC \cite{Aicheler:2012bya, deBlas:2018mhx} that would operate in part as Higgs factories.  

The potential gains of a precision Higgs program pursued at both the LHC and future colliders are innumerable. Confirmation of Standard Model predictions for Higgs properties would mark a triumphant validation of the theory and illuminate phenomena never before seen in nature. The observation of deviations from Standard Model predictions, on the other hand, would point the way directly to additional physics beyond the Standard Model. Such deviations could take the form of changes in Higgs couplings to itself or other Standard Model states, or they could appear as exotic decay modes not predicted by the Standard Model. The latter possibility has been extensively explored for {\it prompt} exotic decay modes in the context of both the LHC (see e.g.~\cite{Curtin:2013fra, deFlorian:2016spz}) and future Higgs factories \cite{Liu:2016zki}. 

However, an equally compelling possibility is for new physics to manifest itself in exotic decays of the Higgs boson to long-lived particles (LLPs). Such signals were first considered in the context of Hidden Valleys \cite{Strassler:2006im, Strassler:2006ri, Han:2007ae} and subsequently found to arise in a variety of motivated scenarios for physics beyond the Standard Model, including solutions to the electroweak hierarchy problem \cite{Craig:2015pha} and models of baryogenesis \cite{Cui:2014twa}; for an excellent recent overview, see \cite{Curtin:2018mvb}. The search for exotic Higgs decays into LLPs necessarily involves strategies outside the scope of typical analyses. The non-standard nature of these signatures raises the compelling possibility of discovering new physics that has been heretofore concealed primarily by the novelty of its appearance.

There is a rich and rapidly growing program of LLP searches at the LHC. A variety of existing searches by the ATLAS, CMS, and LHCb collaborations (e.g.~\cite{CMS:2014hka, Aaboud:2018arf, Aaij:2017mic}; for a recent review see \cite{Lee:2018pag}) constrain Higgs decays into LLPs at roughly the percent level across a range of LLP lifetimes. Significant improvements in sensitivity are possible in future LHC runs with potential improvements in timing \cite{Liu:2018wte}, triggers \cite{Gershtein:2017tsv, CMS:2018qgk, LHCb:2018hiv}, and analysis strategies \cite{Csaki:2015fba, Curtin:2015fna}. Most notable among these is the possible implementation of a track trigger \cite{Gershtein:2017tsv, CMS:2018qgk}, which would significantly lower the trigger threshold for Higgs decays into LLPs and potentially allow sensitivity to branching ratios on the order of $10^{-6}$ in zero-background scenarios. 

While studies of prompt exotic Higgs decays at future colliders \cite{Liu:2016zki} have demonstrated the potential for significantly improved reach over the LHC, comparatively little has been said about the prospects for constraining exotic Higgs decays to long-lived particles at the same facilities.\footnote{A notable exception is CLIC, for which a study of tracker-based searches for Higgs decays to LLPs has been recently performed \cite{Kucharczyk:2625054}. For preliminary studies of other non-Higgs LLP signatures at future lepton colliders, see e.g.~\cite{Antusch:2016vyf}. For studies of LLP signatures at future electron-proton colliders, see \cite{Curtin:2017bxr}.} In this work we take the first steps towards filling this gap by studying the sensitivity of $e^+ e^-$ Higgs factories to hadronically-decaying new particles produced in exotic Higgs decays with decay lengths ranging from microns to meters. For the sake of definiteness we restrict our attention to circular Higgs factories operating at or near the peak rate for the Higgsstrahlung process $\eehz$, namely CEPC and FCC-ee, while also sketching the corresponding sensitivity for the $\sqrt{s} = 250$ GeV stage of the ILC. While essentially all elements of general-purpose detectors may be brought to bear in the search for long-lived particles, the distribution of decay lengths for a given average lifetime make it advantageous to exploit detector elements close to the primary interaction point. We thus focus on signatures that can be identified in the tracker. In order to provide a faithful forecast accounting for realistic acceptance and background discrimination, we employ a realistic (at least at the level of theory forecasting) approach to the reconstruction and isolation of secondary vertices. 

A key question is the extent to which future Higgs factories can improve on the LHC sensitivity to Higgs decays to LLPs, insofar as the number of Higgs bosons produced at the LHC will outstrip that of proposed Higgs factories by more than two orders of magnitude. Higgs decays to LLPs are sufficiently exotic that appropriate trigger and analysis strategies at the LHC should compensate for the higher background rate and messier detector environment. As we will see, there are two natural avenues for improved sensitivity at future lepton colliders: improved vertex resolution potentially increases sensitivity to LLPs with relatively short lifetimes, while lower backgrounds and a cleaner detector environment improves sensitivity to Higgs decays into lighter LLPs whose decay products are collimated. 

The manuscript is organized as follows: In Section \ref{sec:signal} we present a simplified signal model for Higgs decays into pairs of long-lived particles, which in turn travel a macroscopic distance before decaying to quark pairs. We further detail the components of our simulation pipeline and lay out an analysis strategy aimed at eliminating the majority of Standard Model backgrounds. In Section \ref{sec:results} we translate this analysis strategy into the sensitivity of future lepton colliders to long-lived particles produced in Higgs decays as a function of the exotic Higgs branching ratio and the mass and decay length of the LLP. While these forecasts are generally applicable to any model giving rise to the signal topology, we additionally interpret the forecasts in terms of the parameter space of several motivated models in Section \ref{sec:interpretations}. We summarize our conclusions and highlight avenues for future development in \ref{sec:conclusions}.

\section{Signal and Analysis Strategy} \label{sec:signal}

Exotic decays of the Higgs to long-lived particles encompass a wide variety of intermediate and final states. The decay of the Higgs itself into LLPs can proceed through a variety of different topologies. Perhaps the most commonly-studied scenario is the decay of the Higgs to a pair of LLPs, $h \rightarrow XX$, though decays involving additional visible or invisible particles (such as $h \rightarrow X + {\rm invisible}$ or $h \rightarrow XX + {\rm invisible}$) are also possible. The long-lived particles in turn may have a variety of decay modes back to the Standard Model, including $X \rightarrow \gamma \gamma, jj, \ell \bar \ell,$ or $jj \ell,$ including various flavor combinations. These decay modes may also occur in the company of additional invisible states. Moreover, a given long-lived particle may possess a range of competing decay modes, as is the case for LLPs whose decays back to the Standard Model are induced by mixing with the Higgs.

Our aim here is to be representative, rather than comprehensive, as each production and decay mode for a long-lived particle is likely to require a dedicated search strategy. For the purposes of this study, we adopt a simplified signal model in which the Higgs decays to a pair of long-lived scalar particles $X$ of mass $m_X$, which each decay in turn to pairs of quarks at an average ``proper decay length'' $c \tau$.\footnote{Of course, ``proper decay length'' is a bit of a misnomer, but we use it as a proxy for $c$ times the mean proper lifetime $\tau$.} Both the mass $m_X$ and proper decay length $c \tau$ are treated as free parameters, though they may be related in models that give rise to this topology. For the sake of definiteness, for $m_X > 10 \text{ GeV}$ we take a branching ratio of $0.8$ to $b \bar b$ and equal branching ratios of $0.05$ to each of $u \bar u, d \bar d, s \bar s, c \bar c$, though the precise flavor composition is not instrumental to our analysis. For $m_X \leq 10 \text{ GeV}$ we take equal branching ratios into each of the lighter quarks. We further restrict our attention to Higgs factories operating near the peak of the $\eehz$ cross section, for which the dominant production process will be $\eehz$ followed by $h \rightarrow XX$. The associated $Z$ boson provides an additional invaluable handle for background discrimination. Here we develop the conservative approach of focusing on leptonic decays of the $Z$, though added sensitivity may be obtained by incorporating hadronic decays.

Given the signal, there are a variety of possible analysis strategies sensitive to Higgs decays to long-lived particles, exploiting various parts of a general-purpose detector. Tracker-based searches are optimal for decay lengths below one meter, with sensitivity to shorter LLP decay lengths all the way down to the tracker resolution. Timing information using timing layers between the tracker and electromagnetic calorimeter offers optimal coverage for slightly longer decay lengths, while searches for isolated energy deposition in the electromagnetic calorimeter, hadronic calorimeter, and muon chambers provides sensitivity to decay lengths on the order of meters to tens of meters. In principle, instrumenting the exterior of a general-purpose detector with large volumes of scintillator may lend additional sensitivity to even longer lifetimes. In this work we will focus on tracker-based searches at future lepton colliders, as these may be simulated relatively faithfully and ultimately are among the searches likely to achieve zero background while retaining high signal efficiency. 

We define our signal model in \texttt{FeynRules} \cite{Alloul:2013bka} and generate the signal $\eehz \rightarrow XX + \ell \bar \ell $ at $\sqrt{s} = 240$ GeV using \texttt{MadGraph 5} \cite{Alwall:2014hca}. Where appropriate, we will also discuss prospects for Higgs factories operating at $\sqrt{s} = 250$ GeV (potentially with polarized beams) such as the ILC by rescaling rates with the appropriate leading-order cross section ratios. In order to correctly simulate displaced secondary vertices, the decay of the LLP $X$ and all unstable Standard Model particles is then performed in \texttt{Pythia 8} \cite{Sjostrand:2007gs}.

In addition to the signal, we consider some of the leading backgrounds to our signal process and develop selection cuts designed at achieving a zero-background signal region. The most significant irreducible backgrounds from Standard Model processes include $\eehz$ with $Z \rightarrow \ell \bar \ell$ and $h \rightarrow b \bar b$ as well as $e^+ e^- \rightarrow ZZ \rightarrow \ell \bar \ell + j j$. Unsurprisingly, there are a variety of other Standard Model backgrounds, but they are typically well-controlled by imposing basic Higgsstrahlung cuts, and we do not simulate them with high statistics. In addition to irreducible backgrounds from hard collisions, there are possible backgrounds from particles originating away from the interaction point, including cosmic rays, beam halo, and cavern radiation; algorithmic backgrounds originating from effects such as vertex merging or track crossing; and detector noise. Such backgrounds are well beyond the scope of the current study, and will require dedicated investigation with full simulation of the proposed detectors.

Correctly emulating the detector response to LLPs using publicly-available fast simulation tools is notoriously challenging. In particular, we have found that the detector simulator \texttt{Delphes} \cite{deFavereau:2013fsa} tends to cluster calorimeter hits from different secondary vertices into the same jets, significantly complicating the realistic reconstruction of secondary vertices. As such, we develop an analysis strategy using only ingredients from the \texttt{Pythia} output, although we do further run events through \texttt{Delphes} and utilize \texttt{ROOT} \cite{BRUN199781} for analysis. 

\begin{table}[]
    \centering
    \begin{tabular}{|p{6cm}||l|l|}
 \hline
 \text{Cut/Selection} & \text{$ZZ$ Background} & \text{$hZ$ Background} \\
\hline \hline
 \text{Dilepton Invariant Mass} & 0.97 & 0.98 \\
\hline
 \text{Recoil Mass} & 0.006 & 0.94 \\
\hline
 \text{Displaced Cluster ($\geq$ resolution)} & 0.004 & 0.94 \\
\hline
 \text{Invariant Charged Mass (6 GeV)} & 0 & 0.00005 \\
\hline
 \text{Invariant `Dijet' Mass} & 0 & 0.00005 \\
\hline
 \text{Pointer Track} & 0 & 0.00001 \\
\hline
    \end{tabular}
    
    \begin{tabular}
    {|l||p{1.38cm}|p{1.38cm}|p{1.38cm}||p{1.38cm}|p{1.38cm}|p{1.38cm}|}
\hline
$m_X, c \tau$ & $\text{7.5, }10^{-4}$ & $\text{7.5, }10^{-2}$ & $\text{7.5, }10^0$ & $\text{10, }10^{-4}$ & $\text{10, }10^{-2}$ & $\text{10, }10^0$ \\
\hline \hline
 $M_{\ell\ell}$ & 0.97 & 0.97 & 0.97 & 0.97 & 0.98 & 0.97 \\
\hline
 $M_{\text{recoil}}$ & 0.93 & 0.93 & 0.93 & 0.93 & 0.94 & 0.93 \\
\hline
 $|\vec{d}_{\text{cluster}}|$ & 0.93 & 0.93 & 0.41 & 0.93 & 0.94 & 0.50 \\
\hline
 $M_{\text{charged}}$ & 0.27 & 0.28 & 0.08 & 0.55 & 0.55 & 0.21 \\
\hline
  $M_{\text{cluster}}$ & 0.27 & 0.28 & 0.08 & 0.55 & 0.55 & 0.21 \\
\hline
 Pointer & 0.25 & 0.28 & 0.08 & 0.50 & 0.55 & 0.21 \\
\hline
    \end{tabular}
    \begin{tabular}
    {|l||p{1.38cm}|p{1.38cm}|p{1.38cm}||p{1.38cm}|p{1.38cm}|p{1.38cm}|}
\hline
$m_X, c \tau$  & $\text{25, }10^{-4}$ & $\text{25, }10^{-2}$ & $\text{25, }10^0$ & $\text{50, }10^{-4}$ & $\text{50, }10^{-2}$ & $\text{50, }10^0$ \\
\hline \hline
 $M_{\ell\ell}$ & 0.97 & 0.97 & 0.98 & 0.97 & 0.98 & 0.97 \\
\hline
 $M_{\text{recoil}}$ & 0.92 & 0.92 & 0.93 & 0.92 & 0.92 & 0.93 \\
\hline
 $|\vec{d}_{\text{cluster}}|$ & 0.92 & 0.92 & 0.80 & 0.92 & 0.92 & 0.93 \\
\hline
 $M_{\text{charged}}$ & 0.76 & 0.77 & 0.57 & 0.82 & 0.85 & 0.81 \\
\hline
 $M_{\text{cluster}}$ & 0.76 & 0.77 & 0.57 & 0.76 & 0.80 & 0.76 \\
\hline
 Pointer & 0.73 & 0.76 & 0.57 & 0.75 & 0.77 & 0.76 \\
\hline
    \end{tabular}
    \caption{Cut flow of `large mass' analysis for the CEPC with entries of acceptance $\times$ efficiency. The top set of rows gives the cut flow on 500k $Z(jj)Z(\ell \bar \ell)$ events and 100k $h(b \bar b)Z(\ell \bar \ell)$ background events, which are used to confirm our analysis is in the no-background regime. The next sets of rows give cut flows on 5k signal events at representative parameter points, where the different columns are labeled by $m_X/\text{GeV},c\tau/\text{m}$. The full row labels are given in the top set of rows and the labels below are abbreviations for the same cuts or selections. }
    \label{tab:large_mass}
\end{table}

\begin{table}[]
    \centering
    
    \begin{tabular}{|p{6cm}||l|l|}
\hline
 \text{Cut/Selection} & \text{$ZZ$ Background} & \text{$hZ$ Background} \\ 
\hline \hline
 \text{Dilepton Invariant Mass} & 0.97 & 0.98 \\
\hline
 \text{Recoil Mass} & 0.006 & 0.94 \\
\hline
 \text{Displaced Cluster ($\geq 3$ cm)} & 0.004 & 0.62 \\
\hline
 \text{Charged Invariant Mass (2 GeV)} & 0 & 0.002 \\
\hline
 \text{`Dijet' Invariant Mass} & 0 & 0.002 \\
\hline
 \text{Pointer Track} & 0 & 0.001 \\
\hline
 \text{Isolation} & 0 & 0.00005 \\
\hline
    \end{tabular}
    \begin{tabular}{|l||p{1.37cm}|p{1.37cm}|p{1.37cm}||p{1.37cm}|p{1.37cm}|p{1.37cm}|}
    \hline
$m_X, c \tau$ & $\text{2.5, }10^{-4}$ & $\text{2.5, }10^{-2}$ & $\text{2.5, }10^0$ & $\text{7.5, }10^{-4}$ & $\text{7.5, }10^{-2}$ & $\text{7.5, }10^0$ \\
\hline \hline
 $M_{\ell\ell}$ & 0.97 & 0.97 & 0.98 & 0.97 & 0.97 & 0.97 \\
\hline
 $M_{\text{recoil}}$ & 0.93 & 0.93 & 0.93 & 0.93 & 0.93 & 0.93 \\
\hline
$|\vec{d}_{\text{cluster}}|$ & 0.21 & 0.89 & 0.15 & 0.41 & 0.89 & 0.41 \\
\hline
$M_{\text{charged}}$ & 0 & 0.40 & 0.05 & 0 & 0.74 & 0.34 \\
\hline
$M_{\text{cluster}}$ & 0 & 0.40 & 0.05 & 0 & 0.74 & 0.34 \\
\hline
Pointer & 0 & 0.40 & 0.05 & 0 & 0.74 & 0.34 \\
\hline
 Isolation & 0 & 0.33 & 0.045 & 0 & 0.51 & 0.33 \\
\hline

    \end{tabular}

    \begin{tabular}{|l||p{1.37cm}|p{1.37cm}|p{1.37cm}||p{1.37cm}|p{1.37cm}|p{1.37cm}|}
\hline \hline
$m_X, c \tau$ & $\text{15, }10^{-4}$ & $\text{15, }10^{-2}$ & $\text{15, }10^0$ & $\text{50, }10^{-4}$ & $\text{50, }10^{-2}$ & $\text{50, }10^0$ \\
\hline \hline
$M_{\ell\ell}$ & 0.97 & 0.98 & 0.97 & 0.97 & 0.98 & 0.97 \\
\hline
$M_{\text{recoil}}$ & 0.93 & 0.93 & 0.93 & 0.92 & 0.92 & 0.93 \\
\hline
$|\vec{d}_{\text{cluster}}|$ & 0.59 & 0.87 & 0.65 & 0.64 & 0.69 & 0.92 \\
\hline
$M_{\text{charged}}$ & 0.001 & 0.71 & 0.63 & 0 & 0.10 & 0.91 \\
\hline
$M_{\text{cluster}}$ & 0.001 & 0.71 & 0.63 & 0 & 0.09 & 0.90 \\
\hline
Pointer & 0.001 & 0.65 & 0.60 & 0 & 0.08 & 0.84 \\
\hline
Isolation & 0.0002 & 0.42 & 0.58 & 0 & 0.05 & 0.77 \\
\hline
    \end{tabular}

    \caption{Cut flow of `long lifetime' analysis for the CEPC with entries of acceptance $\times$ efficiency. The top set of rows gives the cut flow on 500k $Z(j j)Z(\ell \bar \ell)$ events and 100k $h(b \bar b)Z(\ell \bar \ell)$ background events, which are used to confirm our analysis is in the no-background regime. The next sets of rows give cut flows on 5k signal events at representative parameter points, where the different columns are labeled by $m_X/\text{GeV},c\tau/\text{m}$. The full row labels are given in the top set of rows and the labels below are abbreviations for the same cuts or selections.}
    \label{tab:long_life}
\end{table}

We implement two distinct tracker-based analyses with complementary signal parameter space coverage, which we denote as the `large mass' and `long lifetime' pipelines. We shall eventually see that the former will be effective for $m_X \gtrsim 10 \text{ GeV}$ down to proper decay lengths $c\tau \gtrsim 1 \mu \text{m}$, while the latter is able to push down in $m_X$ by a factor of a few though is only fully effective for $c\tau\gtrsim 1 \text{ cm}$. Full cut tables for both irreducible backgrounds and a variety of representative signal parameter points appear in Tables \ref{tab:large_mass},\ref{tab:long_life} respectively.

As a first step in either analysis, we select Higgsstrahlung events by requiring that our events have an opposite sign electron (muon) pair in the invariant mass range $70 \leq M_{ee} \leq 110$ GeV ($81 \leq M_{\mu\mu} \leq 101$ GeV) and with recoil mass $M_{\text{recoil}}^2 = \big( (\sqrt{s},\vec{0})^\mu - p^\mu_{\ell\ell}\big)^2$ in the range $120 \leq M_{\text{recoil}} \leq 150$ GeV, with $p^\mu_{\ell\ell}$ the momentum of the lepton pair. This allows us to limit our background considerations to the irreducible backgrounds mentioned above and cuts down severely on the $e^+ e^- \rightarrow ZZ$ background, as seen in Tables \ref{tab:large_mass} and \ref{tab:long_life}.

We next identify candidate secondary vertices using a depth-first `clustering' algorithm, which roughly emulates that performed in the CMS search \cite{CMS:2014wda}. We perform this clustering using \textit{all} particles in the event because at later points in the analysis we need this truth-level assignment of neutral particles to clusters, but we expect that this inclusion does not significantly modify the performance of this algorithm. Beginning with a single particle as the `seed' particle for our algorithm, we look through all other particles in the event and create a `cluster' of particles consisting of the seed particle and any others whose origins are within $\ell_\text{cluster} = 7 \ \mu \text{m}$ (the projected tracker resolution of CEPC \cite{CEPCStudyGroup:2018ghi}) of the seed particle. We then add to that cluster any particles whose origins are within $\ell_\text{cluster}$ of any origins of particles in the cluster, and do this step iteratively until no further particles are added to the cluster. We then choose a new seed particle which has not yet been assigned to a cluster and begin this clustering process again. We repeat this process until all particles in the event have been assigned to clusters. We assign to each cluster a location $\vec{d}_\text{cluster}$ which is the average of the origins of all charged particles in the cluster. To ensure that our events contain displaced vertices, we impose a minimum bound on the displacement from the interaction point $|\vec{d}_\text{cluster}| > d_\text{min}$, and clusters satisfying this requirement constitute candidate secondary vertices. For our `large mass' analysis we set $d_\text{min}$ to be the impact parameter resolution ($\simeq 5 \ \mu\text{m}$ for both CEPC and FCC-ee \cite{CEPCStudyGroup:2018ghi},\cite{dam_2017}), and so retain sensitivity to very short $X$ lifetimes. For our `long lifetime' analysis we set $d_\text{min}= 3 \text{ cm}$, which removes the vast majority of clusters coming from $b$ hadron decays in background events, as seen in Table \ref{tab:long_life}. An upper bound $|\vec{d}_\text{cluster}| < r_\text{tracker}$ is imposed by the outer radius of the tracker, where $r_\text{tracker} = 1.81 \text{m}$ for CEPC and $r_\text{tracker} = 2.14 \text{m}$ for FCC-ee are proposed. 

At this point an experimental analysis would want to look at dijets containing candidate secondary vertices and impose an upper bound on the invariant mass to remove dijets coming from $H$ or $Z$ decays. As discussed above we are limited to \texttt{Pythia} objects, but to mock up the (small) penalty to signal of such a selection we implement a selection on the total invariant mass of the clusters $M_{\text{cluster}}=(\sum_{i \in \text{ cluster}} p_i^\mu)^2$. Since this is truth-level information, to turn it into an analog for the dijet invariant mass we apply a Gaussian smearing with a standard deviation of $10 \text{ GeV}$ to account for the dijet resolution.
We then select only candidate secondary vertices with $M_{\text{cluster}} < m_h/2$. This has no effect on background as the background candidate secondary vertices are the result of hadronic decays, so the invariant masses of these clusters are not analogs for dijet invariant masses.

While the total invariant mass of the clusters is not an experimental observable, the invariant mass of charged particles in the clusters $M_{\text{charged}}=(\sum_{i \text{  charged}} p_i^\mu)^2$ is experimentally accessible. For our `large mass' analysis we select candidate secondary vertices with $M_{\text{charged}}> 6 \text{ GeV}$, which gets rid of nearly all clusters from hadronic decays, as seen in Table \ref{tab:large_mass}. For the `long lifetime' analysis, while the increased displacement requirement removes $b$ hadrons it still allows $c,s$ hadrons, and so we select $M_{\text{charged}}> 2 \text{ GeV}$ to address this, which Table \ref{tab:long_life} shows is again very effective.

Next we select the cluster closest to the beamline which passes the above selection requirements as our secondary vertex for the event. Choosing the closest one preferentially selects $X$ decay clusters over hadronic decay clusters in the jets to which the $X$ decays, though this can be fooled by a non-zero fraction of `back-flowing' quarks in $X$ decays (quarks with momenta pointing toward the beamline). 

To remove displaced vertices coming from the decays of charged $b$ hadrons we implement a `pointer track' cut in both analyses as follows. For the cluster selected as the secondary vertex, we consider a sphere of radius $r = 0.5 \text{ mm}$ around the position $\vec{d}_\text{cluster}$. We look for any charged particles whose origins are outside this sphere and whose momenta (at the point at which they were created) point into it, and veto the event if there are any such particles. The main effect of this cut is to remove clusters which were produced from the decay of a charged hadron. The sphere size has been chosen to maximize this effect, though this allows a small effect on signal due to geometric coincidence. Since this cut is only on charged particles, roughly $\sim 30\%$ of background clusters are unaffected. For this cut we ignore the effect of the magnetic field in the tracker, which should not highly impact the trajectories on short scales.

For the `long lifetime' analysis we further implement an `isolation' cut to remove neutral hadronic background decays. Given the cluster selected as the secondary vertex, we consider the plane perpendicular to the sum of momenta of charged particles in the cluster which passes through $\vec{d}_\text{cluster}$. We project the paths of prompt (vertex within $3 \ \mu \text{m}$ of the primary vertex, the planned CEPC vertex resolution \cite{CEPCStudyGroup:2018ghi}) charged particles onto this plane (again ignoring the magnetic field) and veto the event if any come within $R = 10 \text{ cm}$ of the position of the secondary vertex. This radius was chosen to maximally reduce background, and does have a deleterious effect on short decay lengths $\lesssim 10 \text{ mm}$, as can be seen in Table \ref{tab:long_life}. This cut is not perfectly effective at rejecting background due to the non-negligible presence of jets whose prompt components have neutral fraction $1$.
 
\section{Results and Discussion} \label{sec:results}

To confirm that our analysis pipelines put us in the zero-background regime we run both the `long lifetime' and the `large mass' analyses on 500k $e^+ e^- \rightarrow Z(j j)Z(\ell \bar \ell)$ events and 100k $e^+ e^- \rightarrow h(b \bar b) Z(\ell \bar \ell)$ background events. For both pipelines we find that zero $e^+ e^- \rightarrow Z(j j)Z(\ell \bar \ell)$ events remain, while for $e^+ e^- \rightarrow h(b \bar b) Z(\ell \bar \ell)$ we find efficiencies of $5\times 10^{-5}$ and $1\times 10^{-5}$ respectively. We then run each analysis on 5k signal events to get acceptance $\times$ efficiencies for each $(m_X, c\tau)$ point, for a selection of points with   $m_X = 2.5$ from GeV to $50$ GeV and $c\tau$ from $1 \ \mu\text{m}$ to $50$ m. In Table \ref{tab:long_life} we give a cut table for both backgrounds and some representative signal parameter points for the `long lifetime' analysis, and in Table \ref{tab:large_mass} we do the same for the `large mass' analysis.

In the zero-background regime, Poisson statistics rules out model points which predict 3 or more signal events to $95\%$ confidence (or better) if no signal is detected. We may then find a projected $95\%$ upper limit on branching ratio as 
\begin{equation}
    \text{Br}(h\rightarrow XX)^{95} = \frac{N_{sig}}{\mathcal{L}\times \sigma(e^+ e^- \rightarrow hZ) \times \text{Br}(Z\rightarrow \ell \ell) \times A \times \varepsilon},
\end{equation}
with $N_{sig} = 3$ and $A\times \varepsilon$ the result of our simulations. For both the CEPC and FCC-ee, the most recent integrated luminosity projections \cite{CEPCStudyGroup:2018ghi, FCCee} give $\mathcal{L}\times  \sigma(e^+ e^- \rightarrow hZ) = 1.1 \times 10^6$ Higgses produced.

\begin{figure}[t]
  \centering
    \includegraphics[width=\textwidth]{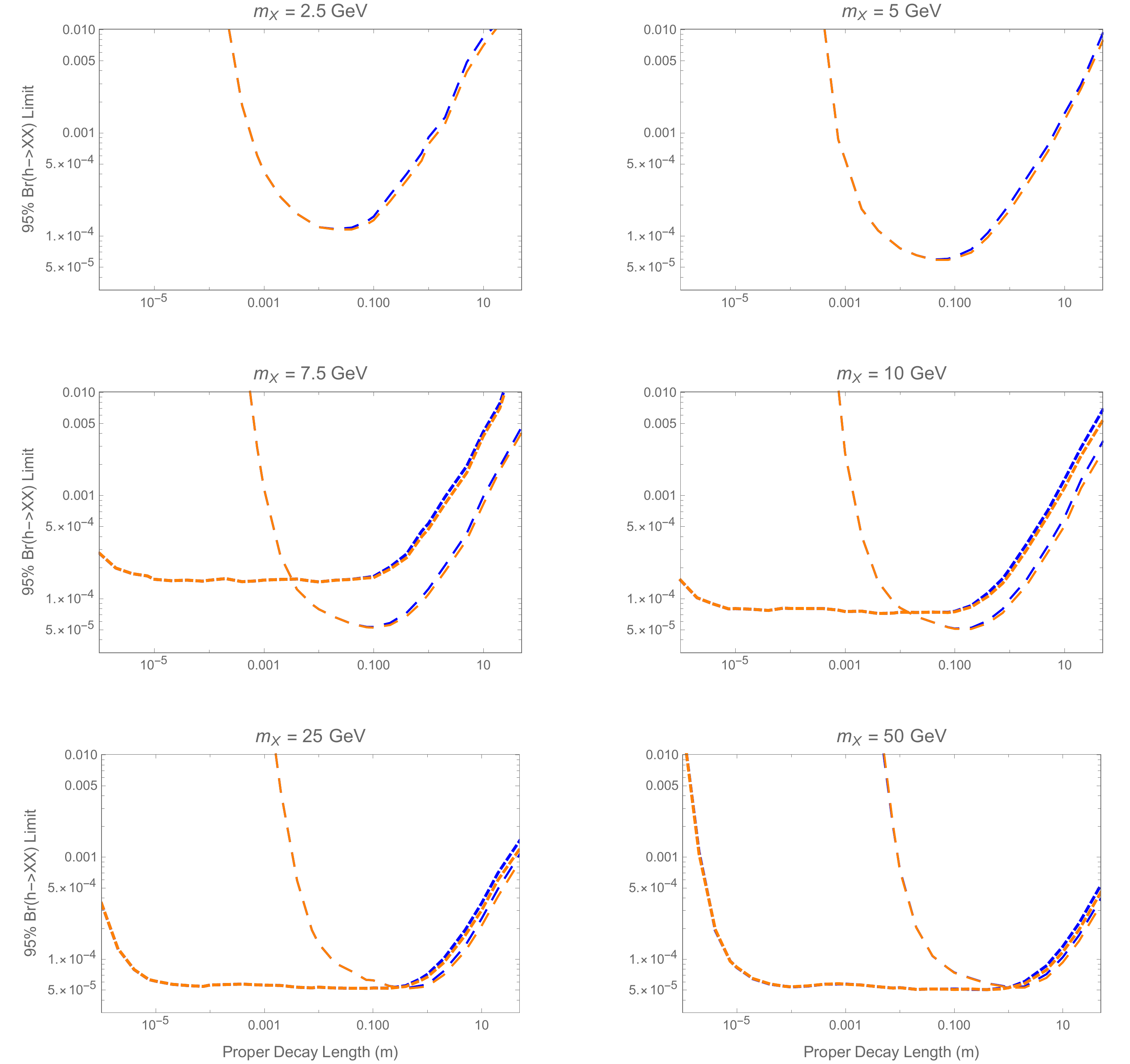}
  \caption{Projected 95\% $h \rightarrow XX$ branching ratio limits as a function of proper decay length for a variety of $X$ masses. Blue lines are for CEPC and orange lines are for FCC-ee, and where only one is visible they overlap. The larger dashes are the `long lifetime' analysis and the smaller dashes are the `large mass' analysis. } \label{fig:brlimits}
\end{figure}

In Figure \ref{fig:brlimits} we show projected 95\% upper limits on $\text{Br}(h\rightarrow XX)$ as a function of $X$ mass and proper decay length. While we plot separate lines for both CEPC and FCC-ee, we only use one set of signal events generated at $\sqrt{s} = 240 \text{ GeV}$ and only account for the difference in tracker radii, so these overlap entirely at smaller lifetimes. 
Approximate limits for the ILC can be obtained by multiplying the above branching ratio limits by a factor of $\sim 1.8$ (i.e.~weakening the limit) to account for the leading order differences in center-of-mass energy, polarization, and integrated luminosity at the $\sqrt{s} = 250$ GeV ILC run, assuming comparable acceptance and efficiency. The ILC limits weaken slightly further for large decay lengths, as its proposed tracker radius is $1.25$ m. Of course, adding the higher-energy ILC runs should significantly improve sensitivity given analyses suitable for the $WW$ fusion production relevant at those energies.

For small masses we are only able to use the `long lifetime' analysis, which requires large displacement from the beamline to cut out the SM $b$ hadron background. As a result we only retain good sensitivity to $X$ decay lengths comparable with the tracker size, though the fact that we only require one displaced vertex (out of two $X$s per signal event) significantly broadens our sensitivity range. This fact also helps us retain efficiency at low masses, as we are able to get down to a projected branching ratio limit of $1\times 10^{-4}$ for $m_X = 2.5$ \text{GeV} despite our $2$ GeV cut on charged invariant mass of the decay cluster. For larger masses this cut has less effect, which allows it to push down to even lower branching ratios $\sim 5 \times 10^{-5}$.

The `large mass' analysis begins working well for masses not far above the $6$ GeV charged invariant mass cut and provides sensitivity to far shorter decay lengths, reaching all the way down to the impact parameter resolution and below. For $m_X = 10$ GeV, where we are aided by the boost factor, we project a limit of $1\times 10^{-4}$ for a proper decay length of $1$ micron. The sensitivity to extremely small decay lengths drops for larger masses, but at $m_X = 50$ GeV we cross below the $10^{-4}$ threshold by $7.5 \ \mu$m. For $X$ masses high enough that the charged invariant mass cut does not remove a large amount of signal events, this analysis projects a branching ratio limit of $\sim 5 \times 10^{-5}$ across roughly the entire range of decay lengths corresponding to the geometric volume of the detector. There is a slight dip in sensitivity for $c\tau \sim 1 \text{ mm}$, where the pair of dijets from the two $X$ decays are most likely to overlap and trigger the cut on `pointer' tracks.

The glaring region of this parameter space to which our analyses do not provide good sensitivity is the low mass $m_X \lesssim 6 \text{ GeV}$ and short proper decay length $c\tau \lesssim 1 \text{ cm}$ regime. The difficulty is that, from the perspective of the tracker, the $X$ here looks more and more like a neutral SM hadron. An analysis making use of the impact parameter distribution of particles in clusters may help here \cite{CMS:2014wda}, but we leave this to future work. Taking advantage of calorimeter data to distinguish between clusters in single jets versus dijets is also likely to provide good sensitivity, but we again leave this to future exploration.

Broadly speaking, our results suggest a peak sensitivity of ${\rm Br}(h \rightarrow XX) \sim 5 \times 10^{-5}$, weakening to $\sim 10^{-4}$ for lower-mass LLPs. Significant additional improvement could be expected with the inclusion of hadronic $Z$ decays, but this requires further study to ensure the control of corresponding Standard Model backgrounds. These limits are competitive with LHC forecasts based on conventional Higgs triggers \cite{Csaki:2015fba, Curtin:2015fna}, noting that these latter forecasts assume zero background. However, the lepton collider limits are potentially superseded by an efficient CMS track trigger \cite{Gershtein:2017tsv, CMS:2018qgk} for higher-mass LLPs, again assuming zero background is achievable with high signal efficiency across a range of lifetimes. In this respect, the primary strengths of the Higgs factories in searching for exotic Higgs decays to LLPs are the potential to push down to shorter decay lengths and lighter LLPs. In particular, the relatively clean and low-background environment of lepton colliders should enable efficient LLP searches even when the LLP decay products become collimated, which remains a weakness of the corresponding LHC searches.

\section{Signal Interpretations} \label{sec:interpretations}

While the bounds presented in the previous section apply to any scenario in which the Higgs decays into pairs of long-lived particles which in turn decay (at least in part) into pairs of quarks,  it is also useful to interpret these bounds in the context of specific models that relate the Higgs branching ratio to LLPs (and the LLP lifetime) to underlying parameters. This illustrates the potential for LLP searches at future lepton colliders to constrain motivated scenarios for physics beyond the Standard Model and allows us to explore the potential complementarity between LLP searches and precision Higgs coupling measurements. To this end, we consider the implications of the LLP limits presented here in the context of both the original Higgs portal Hidden Valley model and a variety of models of neutral naturalness.

\subsection{Higgs Portal}

As a general proxy model for Higgs decays into LLPs, we first consider the archetypal Higgs portal Hidden Valley \cite{Strassler:2006ri}. This entails the extension of the Standard Model by an additional real singlet scalar $\phi$, which couples to the Standard Model through the Higgs portal \cite{Silveira:1985rk, McDonald:1993ex, Burgess:2000yq} via
\begin{equation}
\mathcal{L} \supset \frac{1}{2} (\partial_\mu \phi)^2 - \frac{1}{2} M^2 \phi^2 - A |H|^2 \phi - \frac{1}{2} \kappa |H|^2 \phi^2 -\frac{1}{3!} \mu \phi^3 - \frac{1}{4!} \lambda_\phi \phi^4 - \frac{1}{2} \lambda_H |H|^4.
\end{equation} 
If $\phi$ respects a $\mathbb{Z}_2$ symmetry under which $\phi \rightarrow - \phi$, this additionally sets $\mu = A = 0$, such that the singlet scalar only couples to the Standard Model via the quartic interaction $|H|^2 \phi^2$. After electroweak symmetry breaking, in unitary gauge $H = \left(0, \frac{1}{\sqrt{2}} (h+v) \right)$, but the CP-even scalars $h$ and $\phi$ do not mix. Nonetheless, the quartic interaction nonetheless provides a significant portal for the production of $\phi$, as $\phi$ may be pair produced via the decay $h \rightarrow \phi \phi$ for  $m_\phi < m_h / 2$. Of course, $\phi$ is stable if the $\mathbb{Z}_2$ symmetry is exact, rendering it a potential (albeit highly constrained) dark matter candidate \cite{He:2008qm, Gonderinger:2009jp, Mambrini:2011ik}.

This model gives rise to long-lived particle signatures \cite{Strassler:2006ri} if the $\mathbb{Z}_2$ is broken by a small amount, such that $A \neq 0$ but e.g. $A^2/M^2 \ll \kappa $. The relative smallness of $A$ is technically natural, as the $\mathbb{Z}_2$ symmetry is restored when $A \rightarrow 0$. This then leads to mass mixing between the CP even scalars. As long as $A$ is small compared to $M$ and $v$, the mass eigenstates consist of an SM-like Higgs $h_{\rm SM}$ and a mostly-singlet scalar $s$, related to the gauge eigenstates by 
\begin{eqnarray}
h_{\rm SM} &=& h \cos \theta + \phi \sin \theta \\
s &=& - h \sin \theta + \phi \cos \theta,
\end{eqnarray}
where $\theta \ll 1$ is the mixing angle. There are now two parametrically distinct processes: pair production of the scalar $s$ via Higgs decays, governed by the size of the $\mathbb{Z}_2$-preserving coupling $\kappa$, and decay of the $s$ scalar back to the Standard Model, governed by the size of the $\mathbb{Z}_2$-breaking coupling $A$. In the limit of small mixing, the former process is of order
\begin{equation}
\Gamma(h \rightarrow ss) \approx \frac{\kappa^2 v^2}{32 \pi m_h} \sqrt{1 - 4 \frac{m_s^2}{m_h^2}},
\end{equation}
where we are neglecting subleading corrections proportional to $\lambda_H \sin^2 \theta$. The latter process proceeds into whatever Standard Model states $Y$ are kinematically available, with partial widths
\begin{equation}
\Gamma(s \rightarrow YY) = \sin^2 \theta \times \Gamma (h_{\rm SM}[m_s] \rightarrow YY),
\end{equation}
where $h_{\rm SM}[m_s]$ denotes a Standard Model-like Higgs of mass $m_s$. This naturally leads to a scenario in which the $s$ scalars may be copiously produced via Higgs decays but travel macroscopic distances before decaying back to Standard Model particles. 

This scenario may be constrained not only by direct searches for Higgs decays to LLPs (with the scalar $s$ playing the role of the LLP), but also by precision Higgs coupling measurements. Higgs coupling deviations in this scenario arise from two parametrically distinct effects: tree-level deviations proportional to $\theta^2$ due to Higgs-singlet mixing, and one-loop deviatons proportional to $\kappa$ due to $s$ loops. Both effects result in a universal modification of Higgs couplings, which is best constrained at lepton colliders via the precision measurement of the $\eehz$ cross section \cite{Englert:2013tya, Craig:2013xia}. The net deviation in the $\eehz$ cross section due to these effects in the limit of small mixing is
\begin{equation}
\frac{\delta \sigma_{hZ}}{\sigma_{hZ}^{\rm SM}} \approx - \theta^2 - {\rm Re} \left. \frac{d \mathcal{M}_{hh}}{d p^2} \right|_{p^2 = m_h^2},
\end{equation}
where the radiative correction \cite{Craig:2013xia}
\begin{equation}
 \left. \frac{d \mathcal{M}_{hh}}{d p^2} \right|_{p^2 = m_h^2} = - \frac{1}{16 \pi^2} \frac{\kappa^2 v^2}{2 m_h^2} \left( 1 + \frac{4 m_s^2}{m_h^2} \sqrt{\frac{m_h^2}{m_h^2- 4 m_s^2}} \tanh^{-1} \left[ \sqrt{\frac{m_h^2}{m_h^2 - 4 m_s^2}} \right] \right)
 \end{equation}
 is approximated at $\theta = 0$. Either effect can dominate depending on the relative size of $A/M$ and $\kappa$.

\begin{figure}[t] 
  \centering
    \includegraphics[width=5in]{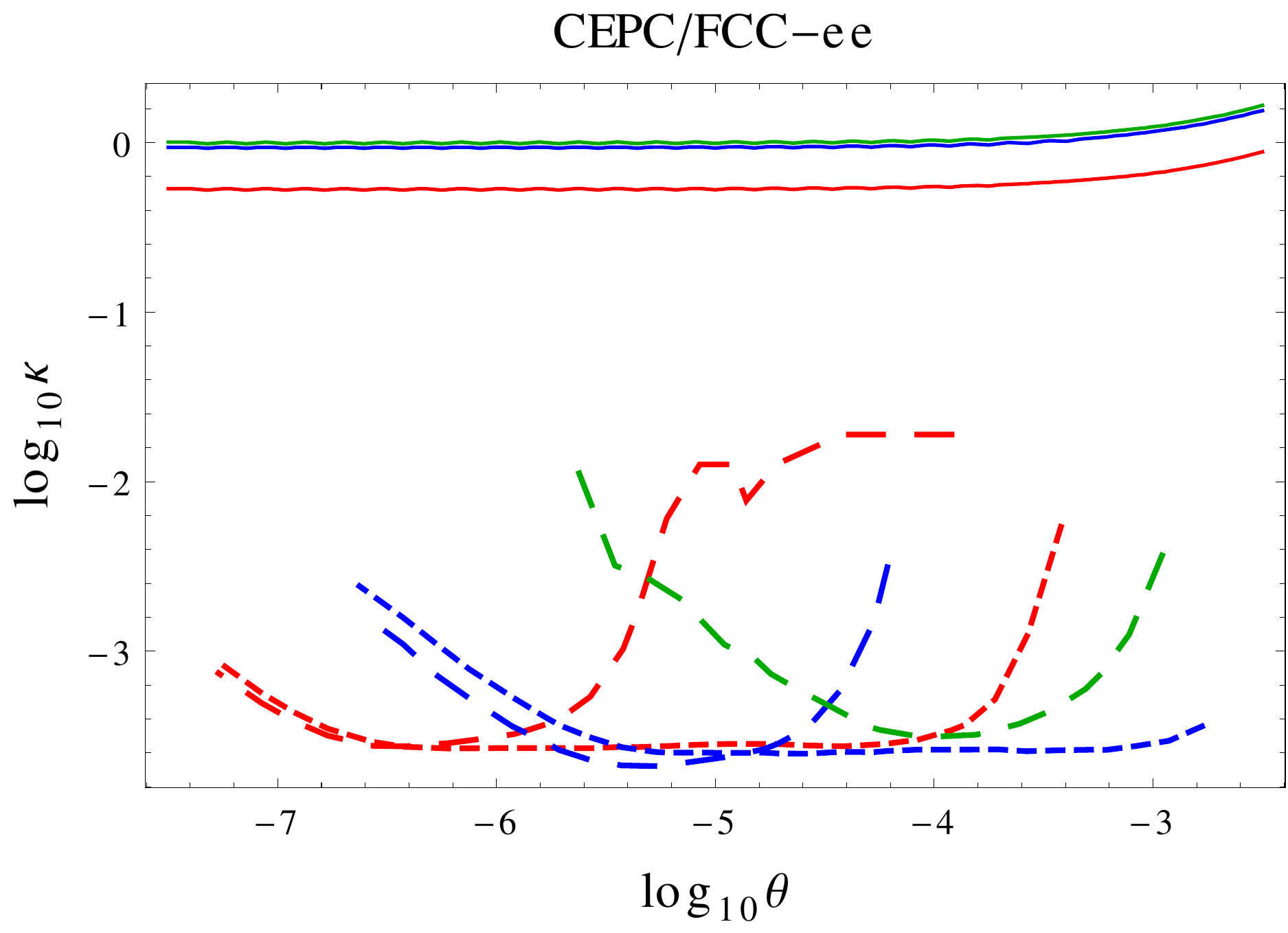}
  \caption{Projected 95\% limits on the Higgs portal Hidden Valley model in the $\kappa, \theta$ plane for three choices of $m_s$; green lines correspond to $m_s = 2.5$ GeV, blue to $m_s = 10$ GeV, and red to $m_s = 50$ GeV.  The solid lines are the projected lower limits from precision Higgs measurements, taking the CEPC projections \cite{CEPCStudyGroup:2018ghi} for definiteness. The dashed lines are projected limits from this work, which are essentially identical for CEPC and FCC-ee.  Long dashes are from the `long lifetime' analysis and short dashes  
  from the `large mass' analysis. } \label{fig:bounds}
\end{figure}

Constraints from a direct search for Higgs decays to LLPs and precision Higgs measurements as a function of the underlying parameters $\theta$ and $\kappa$ are shown in Figure \ref{fig:bounds} for the illustrative benchmarks $m_s = 2.5, 10$, and 50 GeV. Unsurprisingly, in the regime where $s$ is long-lived, the bounds from precision Higgs coupling measurements are modest and direct searches provide the leading sensitivity.

\subsection{Neutral Naturalness}

Higgs decays to LLPs are also motivated by naturalness considerations, arising frequently in models of neutral naturalness that address the hierarchy problem with SM-neutral degrees of freedom \cite{Chacko:2005pe, Craig:2014aea}. In these models, partially or entirely SM-neutral partner particles that couple to the Higgs boson are charged under an additional QCD-like sector. Confinement in the additional QCD-like sector leads to a variety of bound states that couple to the Higgs and may be pair-produced in exotic Higgs decays with predictive branching ratios. The bound states with the same quantum numbers as the physical Higgs scalar typically decay back to the Standard Model by mixing with the Higgs. These decays occur on length scales ranging from microns to kilometers, making them a motivated target for LLP searches at colliders \cite{Craig:2015pha, Curtin:2015fna}. 

For simplicity, here we will restrict our focus to scenarios with the sharpest predictions for the Higgs branching ratio to LLPs. In these cases, the LLPs in question are typically glueballs of the additional QCD-like sector, of which the $J^{PC} = 0^{++}$ is typically the lightest. The coupling of the SM-like Higgs to these LLPs is predominantly due to top partner loops, for which the scales and couplings are directly related to the naturalness of the parameter space. In the Fraternal Twin Higgs \cite{Craig:2015pha}, the entirely SM-neutral fermionic partners of the top quark induce Higgs couplings to twin gluons, which then form glueballs; the $0^{++}$ states are the lightest in the twin QCD spectrum only if the other twin quarks are sufficiently heavy. In addition, there are tree-level deviations in Higgs couplings due to the pseudo-goldstone nature of the SM-like Higgs. In Folded SUSY \cite{Burdman:2006tz}, the scalar top partners carry electroweak quantum numbers, leading to radiative corrections to standard Higgs decays as well as the existence of exotic decay modes. Loops of the scalar top partners again induce Higgs couplings to twin gluons, and without light folded quarks the $0^{++}$ glueball is generically the lightest state in the folded QCD spectrum. While there are no tree-level Higgs coupling deviations in this case, the electroweak quantum numbers of the scalar top partners induce significant corrections to the branching ratio $h \rightarrow \gamma \gamma$. Finally, in the Hyperbolic Higgs \cite{Cohen:2018mgv} (see also \cite{Cheng:2018gvu}), the scalar top partners are entirely SM-neutral, and induce couplings to $0^{++}$ glueballs that are generically the lightest states in the hyperbolic QCD spectrum. As with the Fraternal Twin Higgs, however, there are also tree-level Higgs coupling deviations due to mass mixing among CP-even neutral scalars.

In each of these scenarios, the branching ratio of the SM-like Higgs can be parameterized as follows:
\begin{equation}
{\rm Br}(h \rightarrow 0^{++} 0^{++}) \approx {\rm Br}(h \rightarrow gg)_{\rm SM} \times \left( 2 v^2 \frac{\alpha_s^\prime(m_h)}{\alpha_s(m_h)} \left[ \frac{y^2}{M^2} \right] \right)^2 \times \sqrt{1 - \frac{4 m_0^2}{m_h^2}} 
\end{equation}
Here $\alpha_s^\prime$ denotes the coupling of the additional QCD-like sector (whether twin, folded, or hyperbolic), which is necessarily of the same order as the SM QCD coupling $\alpha_s$, and $m_0$ is the mass of the glueball, which is determined in terms of the QCD-like confinement scale. Adopting the schematic notation of \cite{Curtin:2015fna}, the parameter $\left[ \frac{y^2}{M^2} \right]$ encodes the model-dependence of the Higgs coupling to pairs of gluons in the QCD-like sector, with
\begin{equation}
\left[ \frac{y^2}{M^2} \right] \approx \left \{ \begin{array}{ll} 
- \frac{1}{2v^2} \frac{v^2}{f^2} & {\rm ~Fraternal~Twin~Higgs} \\
\frac{1}{4v^2} \frac{m_t^2}{m_{\tilde t}^2} & {\rm ~Folded~SUSY} \\
\frac{1}{4 v^2} \frac{v}{v_{\mathcal{H}}} \sin \theta & {\rm ~Hyperbolic~Higgs}
\end{array} \right.
\end{equation}
For the Fraternal Twin Higgs, $f$ denotes the overall twin symmetry-breaking scale $f^2 = v^2 + v^{\prime 2} $ in terms of the SM weak scale $v$ and the fraternal weak scale $v^\prime$. For Folded SUSY, $m_{\tilde t}$ denotes the mass of the scalar top partners, neglecting possible mixing effects. For the Hyperbolic Higgs, $v_{\mathcal{H}}$ is the hyperbolic scale and $\tan \theta \approx \frac{v}{v_{\mathcal{H}}}$ encodes tree-level mixing effects. In each case, the scales appearing in the effective coupling are related to the fine-tuning of the model, drawing a direct connection between the Higgs exotic branching ratio and the naturalness of the weak scale.

In each case, the $0^{++}$ glueballs of the additional QCD-like sector decay back to the Standard Model by mixing with the SM-like Higgs, with a partial width to pairs of SM particles $Y$ given by
\begin{equation}
\Gamma(0^{++} \rightarrow YY) = \left( \frac{1}{12 \pi^2} \left[ \frac{y^2}{M^2} \right] \frac{v}{m_h^2 - m_0^2} \right)^2 \left(4 \pi \alpha_s^B F_{0^{++}}^S \right)^2 \Gamma(h_{SM}[m_0] \rightarrow YY),
\end{equation}
where $4 \pi \alpha_s^B F_{0^{++}}^S \approx 2.3 m_0^3$ and, as before, $h_{\rm SM}[m_0]$ denotes a Standard Model-like Higgs of mass $m_0$.

\begin{figure}[t]
  \centering
    \includegraphics[width=\textwidth]{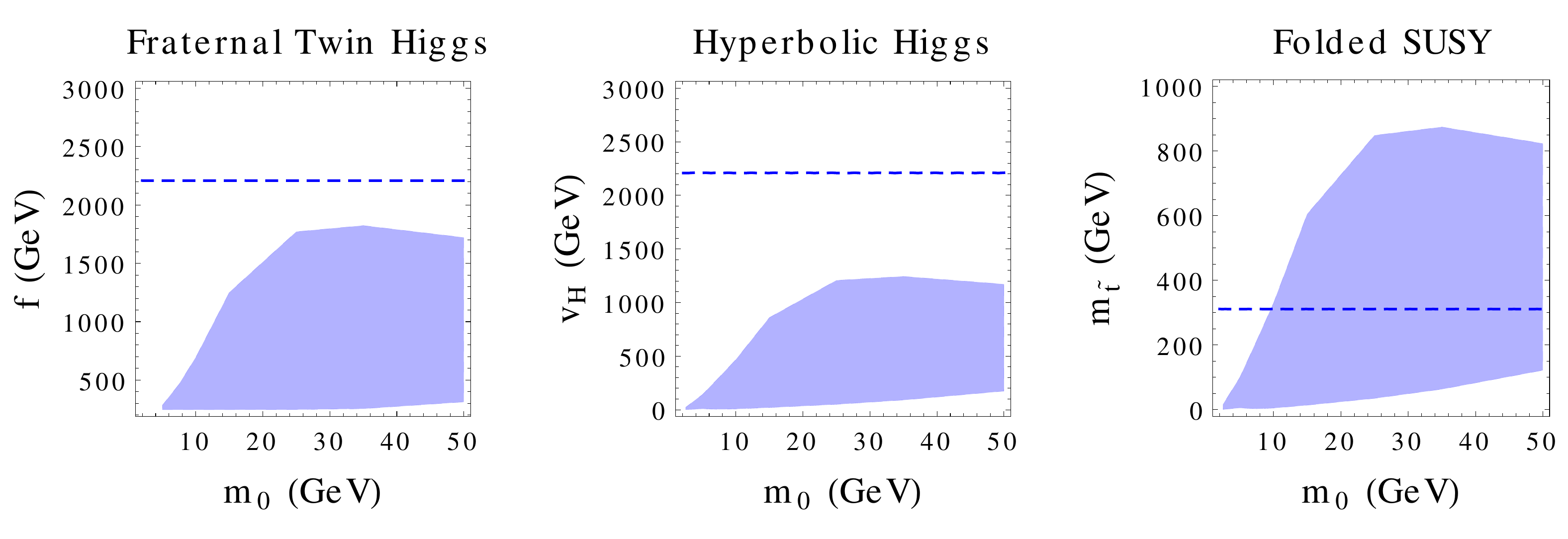}
  \caption{Projected 95\% limits on the underlying scale as a function of the LLP mass $m_0$ in three models of neutral naturalness: the Fraternal Twin Higgs ($f$), the Hyperbolic Higgs ($v_{\mathcal{H}}$), and Folded SUSY ($m_{\tilde t}$). The blue dashed line denotes the limit coming from precision Higgs coupling measurements, taking for definiteness the CEPC projections from \cite{CEPCStudyGroup:2018ghi}. For the Fraternal Twin Higgs and Hyperbolic Higgs, the dominant indirect constraint is from $\sigma_{Zh}$, while for Folded SUSY it is from ${\rm Br}(h \rightarrow \gamma \gamma)$. The shaded region denotes the projected limits from direct LLP searches obtained in this work.} \label{fig:nnplot}
\end{figure}

Constraints on each model from a direct search for Higgs decays to LLPs and precision Higgs measurements are shown in Figure \ref{fig:nnplot} as a function of the LLP mass $m_0$ and the relevant scale ($f, v_\mathcal{H}$, and $m_{\tilde t}$, respectively).   For precision Higgs measurements we use the CEPC projections from \cite{CEPCStudyGroup:2018ghi}. In the Fraternal Twin Higgs and Hyperbolic Higgs, the dominant indirect constraint is from $\sigma_{Zh}$, while for Folded SUSY it is from ${\rm Br}(h \rightarrow \gamma \gamma)$. For both the Fraternal Twin Higgs and the Hyperbolic Higgs, tree-level Higgs coupling deviations make precision Higgs measurements the strongest test of the model. However, the sensitivity of LLPs searches provides valuable complementarity in the event that Higgs coupling measurements yield a discrepancy from Standard Model predictions. In particular, the size of an observed Higgs coupling deviation would single out the relevant overall mass scale ($f$ or $v_{\mathcal{H}}$), providing a firm target for LLP searches that would then validate or falsify these models as an explanation of the deviation. Note also that in the Fraternal Twin Higgs there may be additional contributions to the Higgs branching ratio into LLPs coming from the production of twin bottom quarks, which could lead to sensitivity in the LLP search comparable to that of Higgs couplings.  In the case of Folded SUSY, the absence of tree-level Higgs coupling deviations and the relatively weaker constraints on ${\rm Br}(h \rightarrow \gamma \gamma)$ make the LLP search the leading test of this model at Higgs factories.

\section{Conclusion} \label{sec:conclusions}

The exploration of exotic Higgs decays is an integral part of the physics motivation for future lepton colliders. New states produced in these exotic Higgs decays may themselves decay on a variety of length scales, necessitating a range of search strategies. While considerable attention has been devoted to the reach of future lepton colliders for promptly-decaying states produced in exotic Higgs decays, the reach for long-lived particles is relatively unexplored. 

In this paper we have made a first attempt to study the reach of proposed circular Higgs factories such as CEPC and FCC-ee (as well as approximate statements for the $\sqrt{s} = 250$ GeV run of the ILC) for long-lived particles produced in exotic Higgs decays, focusing on the pair production of LLPs and their subsequent decay to pairs of quarks. We have developed a realistic tracker-based search strategy motivated by existing LHC searches that entails the reconstruction of displaced secondary vertices. Rather than relying on existing public fast simulation tools, which do not necessarily give a sensible parameterization of signal and background efficiencies for long-lived particle searches, we have implemented a realistic approach to clustering and isolation. This allows us to characterize some of the leading irreducible Standard Model backgrounds to our search and determine reasonable analysis cuts necessary for a zero-background analysis. We obtain forecasts for the potential reach of CEPC and FCC-ee on the Higgs branching ratio to long-lived particles with a range of lifetimes. The projected reach is competitive with LHC forecasts and potentially superior for lower LLP masses and shorter lifetimes. In addition to our branching ratio limits, which may be freely interpreted in a variety of model frameworks, we interpret our results in the parameter space of a Higgs portal Hidden Valley and various incarnations of neutral naturalness, demonstrating the complementarity between direct searches for LLPs and precision Higgs coupling measurements.

There are a variety of directions for future work. While we have attempted to investigate some of the leading irreducible backgrounds and impose realistic cuts, we have not attempted to estimate possible backgrounds coming from cosmic rays; algorithmic, detector, or beam effects; or other contributions. Our tracker-based analysis has focused on Higgs decays to pairs of hadronically-decaying LLPs, but a comprehensive picture of exotic Higgs decays would also suggest the investigation of Higgs decays to various LLP combinations as well as the consideration of additional LLP decay modes.  Moreover, tracker-based searches for displaced vertices are but one of many possible avenues to discover long-lived particles. Analogous searches based on timing or on isolated energy deposition in outer layers of the detector (including either the electromagnetic or hadronic calorimeter, the muon chambers, or potentially instrumented volumes outside of the main detector) would be valuable for building a complete picture of LLP sensitivity across a range of lifetimes. 

More broadly, it is an ideal time to study the potential sensitivity of future Higgs factories to long-lived particles, as the results are likely to inform the design of detectors for these proposed colliders. This is a necessary step in motivating the physics case of future Higgs factories and ensuring that they enjoy optimal coverage of possible physics beyond the Standard Model.

\section*{Acknowledgments}
We thank Nick Amin, James Beacham, Stefania Gori, Simon Knapen, Eric Kuflik, Zhen Liu, Bennett Marsh, Simone Pagan Griso, Diego Redigolo, and Sicheng Wang for useful conversations. This research is supported in part by the US Department of Energy under the Early Career Award DE-SC0014129 and the Cottrell Scholar Program through the Research Corporation for Science Advancement. 

\bibliography{LLPBib}

\end{document}